\documentclass[journal,a4paper]{IEEEtran}

\usepackage{graphicx} 
\usepackage{matrices}
\usepackage{custom}

\begin{document}

%\journal{IEEE Trans. Signal Processing}

\title{Random Paraunitary Projections}

\author{
Ricardo L. de Queiroz 
\thanks{The author is with the Department of Computer Science, Universidade de Brasilia, Brazil, e-mail queiroz@ieee.org.} 
}

\markboth{Preprint 2011}{Lab Notes 2011.}

\maketitle

%   ABSTRACT 

\begin{abstract}
Transforms using random matrices have been found to have many applications. 
We are concerned with the projection of a signal onto Gaussian-distributed random orthogonal bases. 
We also would like to easily invert the process through transposes in order to facilitate iterative reconstruction. 
We derive an efficient method to implement random unitary matrices of larger sizes through a set of Givens rotations. 
Random angles are hierarchically generated on-the-fly and the inverse merely requires traversing the angles in reverse order.
Hierarchical randomization of angles also enables reduced storage.
Using the random unitary matrices as building blocks we introduce random paraunitary systems (filter banks).
We also highlight an efficient implementation of the paraunitary system and of its inverse. 
We also derive an adaptive under-decimated system, wherein one can control and adapt the amount of projections the signal undergoes, in effect, varying the sampling compression ratio as we go along the signal, without segmenting it. 
It may locally range from very compressive sampling matrices to (para) unitary random ones. 
One idea is to adapt to local sparseness characteristics of non-stationary signals.
\end{abstract}

% INTRODUCTION

\section{Introduction}

Random matrices \cite{mehta1990} are very popular in multivariate statistics and have  been used in number theory,  nuclear physics, quantum information, mechanics, and wireless telecommunications.
More recently, it has also been used in compressive sensing \cite{cs-baraniuk}, wherein a signal, assumed sparse in some domain, is projected onto a random matrix of reduced dimensionality.
Easily computing the inverse or transpose of the projection may be useful for iterative reconstruction systems such as COSAMP \cite{cosamp}. 
We are interested in an algorithm that would easily allow transforming very large vectors using Gaussian-distributed random orthogonal transforms or filter banks that could be easily reversed. 
The transform should be randomly generated on-the-fly rather than pre-stored.

\section{Random unitary transform}

\label{sec:unitary}

A unitary matrix can be decomposed into plane rotations, known as Givens rotations \cite{hohn-algebra}. 
An $M\times M$ unitary matrix can be decomposed into $M(M-1)/2$ rotations, i.e. rotations involving every pair of axis of the transformation. 
Let $\bR_{ij}$ be a matrix with elements $\{r_{ij}\}$ representing the rotation of an angle $\theta_{ij}$ along the plane containing the $i$-th and $j$-th axes of the transform, i.e. 
$\bR_{ij}$ is like the identity matrix but replacing the elements 
$r_{ii} = r_{jj} = \cos(\theta_{ij})$ and $r_{ji} = -r_{ij} = \sin(\theta_{ij})$.
Let \bS\ be a reflection matrix, i.e. a diagonal matrix with $\pm 1$ in its diagonal. 
Then, the unitary transformation \bU\ can also be represented as 

\beq
\bU = \bS \prod_{i=0}^{M-2} \prod_{j=i+1}^{M-1} \bR_{ij} ,
\label{eq:givens}
\enq

\noindent 
which is the expression for the Givens factorization \cite{hohn-algebra}. 
The inverse transform is easily found by reversing the order of the rotations and inverting the angles. 

Unitary random matrices have been generated by QR factorization of non-unitary random matrices \cite{mezzadri2007}, and by using Householder rotations \cite{stewart1980}.  
We generate unitary random matrices by randomly generating the rotation angles $\theta_{ij}$ rather than randomly generating the matrix elements $u_{ij}$ \cite{anderson1987}.  
The random matrix can be computed using \eqref{eq:givens}, so that the resulting matrix \bU\ is perfectly unitary.
Nevertheless, the statistical correlation and probability distribution function (PDF) of the samples of \bU\ are not  trivially found from the PDF of $\theta_{ij}$ \cite{anderson1987},\cite{heiss1994}. 
If $\theta$ is uniformly distributed, the PDFs of $\cos(\theta)$ and of $\sin(\theta)$ are biased and the many operations derived from the successive plane rotations tend to further concentrate the samples.
Under certain restrictions it can be shown that \cite{heiss1994} if each angle is randomly chosen according to the following ditribution:

\beq
p(\theta_{ij}) = \frac{ \Gamma \left(\frac{j - i + 1}{2}\right) } { \sqrt{\pi} \ \Gamma\left(\frac{j-i}{2}\right)}
\cos^{j-i-1}(\theta_{ij}) ,
\label{eq:heiss}
\enq

\noindent 
then the resulting PDF would tend to be Gaussian. 
Also, of great importance to us is the autocorrelation of the resulting matrix, which ideally should be an impulse.  
The trivial approach is to generate random matrix entries in order to ensure decorrelation. 
Even though we use an indirect method, generating random rotation angles, our simulations show that the above distribution of angles leads to decorrelated near-perfect Gaussian-distributed entries.

It is a case of interest to project an $M$-tuple \bx\ into $N << M$ bases, i.e. an orthogonal $N\times M$ Gaussian random transform \bA, so that $\by = \bA\bx$.
This can be easily accomplished through pruning the unnecessary rotations. 
One can obtain \bA\ through generating \bU\ and discarding $M-N$ rows.
For that, it can be shown that we may also discard the last $M-N-1$ stages in \eqref{eq:givens}.
Its transpose (rather than its inverse transform) $\hat \bx = \bA^T \by$, requires traversing the rotations backwards from $N$ samples to an approximation of the original $M$-tuple.

\subsection*{Low-memory implementation}

If we implement the transform through in-place rotations, we can completely avoid storing or calculating the matrix entries. 
Angles are generated as they are used in sequence following \eqref{eq:givens}.
Each stage (rotation) demands calculating $\cos(\theta_{ij})$ and $\sin(\theta_{ij})$, 4 multiplications and 2 additions.
There is no need for matrix inversion and we only store the input vector. 
The inverse transform is accomplished by reversing the order of the planes and by applying negative rotations. 
However, the angles would have to be generated in reverse order which normally would require to buffer all the random angles. 
Let the set of $\ell=M(M-1)/2$ angles be $\bTheta = \{\theta_0, \ldots , \theta_{\ell-1}\}$. 
We divide the set of angles into $N_s$ subsets $\bTheta_i$ of $\ell/N_s$ angles each as $\bTheta = \{ \bTheta_1 , \bTheta_2, \ldots , \bTheta_{N_s}\}$. 
We first generate a seed $S_0$, from which we generate $N_s$ seeds $S_k$, $1\leq k \leq N_s$. 
Each of the subsets $\bTheta_k$ is then randomly generated using seed $S_k$. 
For the inverse transform angles and seeds, one would need to convey $S_0$ to both forward and inverse transform stages and to buffer only the set of seeds $\{S_k\}$.
At a time, only $\ell/N_s$ angles for one subset need to be generated and visited in reverse order.   
Thus, the only storage required would be necessary for 3 vectors: input/output, seeds, and random subgroup, 
of sizes, $M$, $N_s$ and $\ell/N_s$, respectively.  
If $N_s=M$, then total storage is roughly $2.5 M$. 
This makes it possible to make random orthogonal projections (and their inverses) of very large vectors. 

As a note, we disregarded the computation and storage to produce the random angles with PDF given by \eqref{eq:heiss}.

\section{Random Paraunitary Systems} 

An $M$-input, $M$-output ($M\times M$) paraunitary system $\bH(z)$ of order\footnote{We mean the order of the system as the highest polynomial degree of its entries, and not the McMillan degree of the system \cite{ppbook}.} $K$ can be constructed using a cascade of $K$ simpler order-1 paraunitary systems: 

\beq
\bH(z) = \bU_0 \prod_{i=1}^{K} \bLambda(z) \bU_{i} ,
\label{eq:paraunitary}
\enq

\noindent 
where $\bLambda(z) = diag \{1, \ldots, 1, z^{-1}, \ldots , z^{-1}\}$ and $\bU_i$ are unitary matrices. 

We can make a random, Gaussian, paraunitary system if we cascade random unitary matrices, i.e. if we make $\bU_i$ random unitary matrices as explained in the previous Section. 
To see that, we begin by showing that $\bH(z)$ has random independent entries $\{h_{ij}(z)\}$. Let $\bH(z)=\bH^\prime(z)\bU_n$ such that $h_{ij}(z) = \sum_k h^\prime_{ik}(z)u_{n,kj}$ and that the $\{u_{n,ij}\}$ have zero mean. Then, 
$E\{h_{ij}(z)h_{k\ell}(z)\}=\sum_{uv} E\{ h^\prime_{uj}(z) h^\prime_{v\ell}(z)\} E\{ u_{0,iu} u_{0,kv}\} = 0$ if $i,j \neq k,\ell$ and if the entries $\{h^\prime_{ij}\}$ are independent. 
Since $\bU_0$ has independent entries, so does $\bU_0\bLambda(z)$ and, hence, the same applies to $\bH(z)$ for any order. 

Let $\bF(z)=\sum_{i=0}^{n} \bF_i z^{-i}$ be a system of order $n$, obtained through  appending an order-1 stage to 
$\bE(z)=\sum_{i=0}^{n-1} \bE_i z^{-i}$, i.e. $\bF(z)=\bE(z)\bLambda(z)\bU_n$. 
Let $\bLambda(z)\bU_n=\bU^{\prime}_n+z^{-1}\bU^{\prime\prime}_n$. Then, we get

\[
\bF_k=\bE_k\bU^\prime_n + \bE_{k-1}\bU_n^{\prime\prime} \ \ \ 0 < k < n
\]
\[\bF_0=\bE_0\bU^\prime_n \ \ \  \bF_n=\bE_{n-1}\bU^{\prime\prime}_n . 
\]

The entries $\{f_{k,ij}\}$ of $\bF(z)$ are sums of products of random variables. 
For large $M$, it is expected that the distribution of $\{f_{k,ij}\}$ approaches a Gaussian PDF.  
As we can make all $\bU_i$ to have Gaussian-distributed entries as in Section \ref{sec:unitary}, we expect all $\{h_{k,ij}\}$entries in $\bH(z)=\sum_{i=0}^{K} \bH_i z^{-i}$ to be approximately Gaussian distributed, for large $M$. 

Let $\bA^H$ denote the Hermitian of matrix $\bA$, i.e. its transposed conjugate.
Also, let the subscript $\bA^{(*)}(z)$ denote to conjugate only the coefficients of its polynomials, i.e. if $\bA(z)$ has entries 
$\{\sum_n a_{ijn} z^n\}$, then $\bA^{(*)}(z)$ has entries $\{\sum_n a^*_{ijn} z^n\}$.
Then, the inverse of a paraunitary system $\bH(z)$ is simply its para-conjugated version \cite{ppbook}

\[
\bH^{-1}(z) = \bH^{T(*)}(1/z) = \sum_{i=0}^{K} \bH_i^H z^{i} .
\]

A paraunitary $\bH(z)$ is, thus, unitary on the unit circle. 

\subsection*{Low-memory implementation} 
The system can be easily implemented using \eqref{eq:paraunitary}, wherein each random unitary stage $\bU_i$ can be implemented using the techniques described in the previous section. 
Each $M$-sample vector of the input signal undergoes a chain of random transforms, for each $\bU_i$ interspersed with shuffles $\bLambda(z)$. 
One has to budget memory to store the $KM/2$ internal states, in between the random unitary transforms. 

The inverse transform can be accomplished by traversing the system backwards. Let $\tilde\bLambda(z)=z^{-1}\bLambda(1/z)$, such that $\tilde\bLambda(z)\bLambda(z)=z^{-1}\bI$. Let 

\beq
\bG(z) =  \bU^H_K \prod_{i=1}^{K} \tilde\bLambda(z)\bU^H_{K-i}  .
\label{eq:inverse}
\enq

\noindent 
Then, $\bG(z)\bH(z)=z^{K}\bI$ and the above inverse has the same structure as \eqref{eq:paraunitary}.
Hence, both the forward and inverse transforms have the same implementation, with the proper adaptation of the shuffles in between stages of unitary transforms. 
The transpose (inverse) of a unitary transform can be accomplished by going through the plane rotations in reverse order, with inverse angles.  

As a note, the order can be changed on-the-fly by adding or removing stages to the cascade of unitary transforms.

\section{Adaptive sampling matrices} 

The $M\times M$ system $\bH(z)$ can be seen as an $M$-channel filter bank, where each filter is down-sampled  by a factor of $M$, ($\downarrow M$).
The filters have length $L = (K+1)M$.
In such, $M$ samples enter the system and $M$ samples leave it at a time, in what is referred as a critically decimated system.
If we increase the decimation, the systems becomes over-decimated, and not invertible.
Let us denote as $\bH_{\downarrow m}(z)$ the representation of the paraunitary system, but with down-sampling by a factor of $m$. 
Then, 

\begin{equation}
\bH_{\downarrow M}(z) = \bH(z) = \sum_{i=0}^{K} \bH_i z^{-i}
\end{equation}
\begin{equation}
\bH_{\downarrow 2M}(z) = \sum_{i=0}^{\lfloor K/2\rfloor} \left[ \bH_{2i} , \bH_{2i+1}\right] z^{-i}
\end{equation}

\noindent 
where $\lfloor\ \rfloor$ is the ``floor'' rounding operation. For a more general down-sampling factor, 

\begin{equation}
\bH_{\downarrow qM}(z) = \sum_{i=0}^{\lfloor K/q\rfloor} \left[ \bH_{qi} , \bH_{qi+1}, \ldots, \bH_{qi+q-1}\right] z^{-i} , 
\end{equation}

\noindent where $\bH_n=\bf{0}$ for $n > K$. 
In the extreme case, if $q = K+1$, i.e. $\downarrow L$, then 

\begin{equation}
\bH_{\downarrow L}(z) = \left[ \bH_0, \bH_1, \ldots , \bH_K \right] ,
\end{equation}

\noindent
i.e. the under-sampled matrix becomes a (scalar) sampling matrix. 

There are $qM$ input samples per $M$ output ones, i.e. $q$ is the sampling compression parameter. 
The down-sampling and, thus, $q$ can be changed on the fly. 
The system can go from $M$ input to $M$ output samples per clock, to an $L$-to-$M$ one, and all stages in between. 
In all cases, the bases are random in nature and are approximately Gaussian distributed as discussed in the previous Section. 
Hence, $q$ can be made adaptive, perhaps adapting how compressive the {\em sensing} is, in order to track local sparseness of non-stationary signals.  

\subsection*{Low-memory implementation} 
\begin{figure}
\centering

\includegraphics[width=0.9\columnwidth]{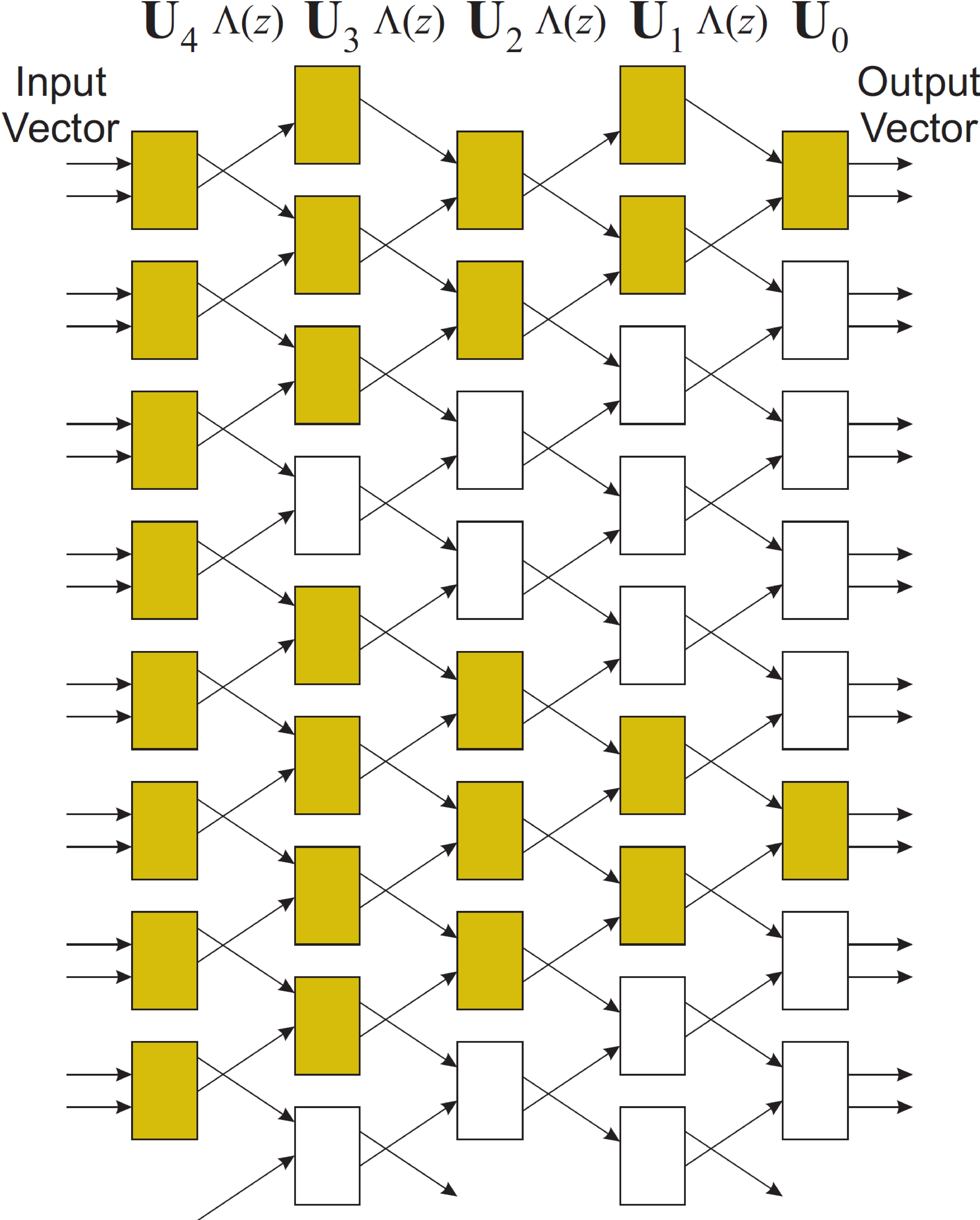}

\caption{A $K=4$ paraunitary system lattice and diagram illustrating the blocks ignored in the $\downarrow L$ case (white filled). Each branch carries $M/2$ samples.}
\label{fig:graph}
\end{figure}

Starting from a $\downarrow M$ system in \eqref{eq:paraunitary}, a $\downarrow qM$ system can be implemented by picking one out of every set of $q$ output blocks of $M$ samples. 
If we view \eqref{eq:paraunitary} as a flow graph, the system can then be more efficiently implemented by simply pruning out the stages which are not necessary and will not be used to produce output.
Figure \ref{fig:graph} illustrates a case $K=4$ and indicates unitary stages to be pruned in the $\downarrow L$ (sampling matrix) case.  
It can be shown that maximum pruning yields computational savings in the order of $K-(2K)^{-1}$.
The reverse (transpose) system can be accomplished through reversing the system, i.e. using \eqref{eq:inverse} and setting to zero the samples which are not produced using the over-decimated system.

%%%%%%%%%%%%%%%%%%%%%%%%%%%%%%%%%%%%%%%%%%%%%%%
%%%%%%%%%%%%%%%%%%%%%%%%%%%%%%%%%%%%%%%%%%%%%%%

\section{Conclusions}

In this letter, we present a method to construct Gaussian-distributed random paraunitary systems.
The system can be dynamically applied in order to create an adaptive compressive sensing framework wherein sampling is compressed by a factor of $q$ which can be changed on-the-fly.
The bases are unitary in nature and the projection can be easily implemented and reversed.  
We discussed the efficient implementation of random unitary matrices, which are used as building blocks to construct the random paraunitary filter banks.
An efficient implementation of such filter banks is also discussed. 
Applications of such an adaptive system are being investigated.

\end{document}